\documentclass[aps,pre,floats,twocolumn,showpacs,superscriptaddress]{revtex4}

\usepackage{graphicx,epsfig}
\usepackage{times}
\usepackage{graphics,dcolumn,bm,fleqn,float}
\usepackage{amssymb,amsmath,multirow,rotate,color}
\begin{document}

\title{Inter-groups information exchange drives Cooperation in the Public Goods Game}

\author{C. Gracia-L\'azaro}

\affiliation{Institute for Biocomputation and Physics of Complex
Systems (BIFI), University of Zaragoza, Zaragoza 50009, Spain}

\author{J. G\'omez-Garde\~nes}

\affiliation{Institute for Biocomputation and Physics of Complex
Systems (BIFI), University of Zaragoza, Zaragoza 50009, Spain}

\affiliation{Departamento de F\'{\i}sica de la Materia Condensada,
University of Zaragoza, Zaragoza E-50009, Spain}

\author{L.M. Flor\'{\i}a}

\email{mario.floria@gmail.com}

\affiliation{Institute for Biocomputation and Physics of Complex
Systems (BIFI), University of Zaragoza, Zaragoza 50009, Spain}

\affiliation{Departamento de F\'{\i}sica de la Materia Condensada,
University of Zaragoza, Zaragoza E-50009, Spain}

\author{Y. Moreno}

\affiliation{Institute for Biocomputation and Physics of Complex
Systems (BIFI), University of Zaragoza, Zaragoza 50009, Spain}

\affiliation{Departamento de F\'{\i}sica Te\'orica,
University of Zaragoza, Zaragoza E-50009, Spain}

\affiliation{Complex Networks and Systems Lagrange Lab, Institute for Scientific Interchange, Turin, Italy}

\date{\today}

\begin{abstract}
In this manuscript we explore the onset of cooperative traits in the Public Goods game. This well-known game involves $N$-agent interactions and thus reproduces a large number of social scenarios in which cooperation appears to be essential. Many studies have recently addressed how the structure of the interaction patterns influences the emergence of cooperation. Here we study how information about the payoffs collected by each individual in the different groups it participates in, influences the decisions made by its group partners. Our results point out that cross-information plays a fundamental and positive role in the evolution of cooperation for different versions of the Public Goods game and different interaction structures. 

\end{abstract}

\pacs{89.75.Fb, 05.45.-a, 05.10.Gg}

\maketitle

\section{Introduction}
\label{Introduction}

While fundamental interactions of matter are of a pairwise nature, this is not, in general, the case 
for interactions among constituents of biological, social or economical systems, where $N$-agent 
interactions could be as fundamental as two-agent ones. Computational approaches aimed at modeling 
the dynamical aspects of these complex systems have traditionally paid much more attention to 
pairwise interactions, but this could be in many instances an oversimplifying assumption, in the 
extent that in general, group interactions are not reducible to the aggregate of two-body interactions. 
For a wide variety of issues and important questions in these kinds of complex systems, a very successful 
formulation of the system's dynamics is that of evolutionary game dynamics 
\cite{Hofbauer,Gintis,Nowak_book,Sigmund}, where interaction among 
agents is modeled as a game, with different possible strategies, from which agents receive payoffs 
and strategies spread over the population in proportion to the payoff obtained, so implementing the 
(Darwinian) natural selection of strategies. 

In this paper we are concerned with the evolutionary game dynamics of a particularly important (due 
to its applications to biological \cite{Chuang,Gore,Frey,Cremer} and socio-economical \cite{Kagel} 
systems) representative of the class of group interactions: the public goods game (PGG). In this 
game, the agents can adopt one of two strategies: Cooperation or Defection. Cooperators (also termed as 
producers) contribute at their cost to the benefits of all members of the group, while defectors (free riders) 
do not contribute to the group welfare, but they enjoy benefits. In a general setting of the PGG, both the cost 
paid by a cooperator, $\zeta(\rho_C)$, and the benefit received by a group member, $\beta(\rho_C)$, are 
arbitrary functions of the fraction, $0\leq\rho_C\leq1$, of cooperators in the group, so that the net benefit 
of a cooperator is $P_C = \beta(\rho_C) - \zeta(\rho_C)$, while that of a free rider is $P_D = \beta(\rho_C)$. 
In the realm of classical game theory ({\em i.e.,} the analysis of decision-making by a rational agent), for a 
constant individual cooperation cost, $\zeta$, and arbitrary convex (or linear) benefit function, $\beta(\rho_C)$, 
free riding is the rational choice, though for concave functions $\beta$ there are intervals of $\rho_C$ values 
for which the corresponding group composition is a Nash equilibrium \cite{Montro}. The most used payoff functions in the literature, which are the ones that will be considered hereafter, are a constant cost 
$\zeta$ and a linear benefit $\beta= r \rho_C$ where $r>0$ is called the {\em synergy} or {\em enhancement} 
factor.

The simplest implementation of evolutionary dynamics (well-mixed approximation) for the PGG (see for example 
\cite{Archetti-Scheuring_12,Archetti-Scheuring_11,Cressman}) assumes that the $N-1$ agents that form a group 
with the focal player are randomly sampled {\em without assortment} from an infinite population of cooperators 
and defectors with an instantaneous fraction $x=\langle\rho_C\rangle$ of cooperators. In this way, the probability 
that the focal player interacts with $j$ cooperators is $f_j(x) = C_j^{N-1}x^j (1-x)^{N-1-j}$ and thus the expected 
payoff of a defector focal player is: 
\begin{equation}
W_D(x) =N^{-1} \sum_{j=0}^{N-1} f_j(x) rj\zeta \;, 
\end{equation}
while for a cooperator focal player it reads:
\begin{equation} 
W_C(x) = W_D(x) +r\zeta N^{-1}-\zeta \;,
\end{equation}
that leads to the replicator equation for the evolution of the expected value $x$ of the fraction of 
cooperators:
\begin{equation}
\dot{x} = x(1-x)[W_C(x) - W_D(x)]\;.
\label{replicator}
\end{equation}
From the $W_C$ and $W_D$ expressions above, one can easily see that for $r/N <1$ cooperation is asymptotically fixed in the population, while if $r/N>1$ defection dominates. 

In order to go beyond the likely unrealistic assumption of random grouping without any assortment, one can place 
agents on the nodes of a graph (network) \cite{Brandt,Hauert-Szabo,Santos08,Szolnocki09,Szolnocki2011} and 
consider that groups in which the PGG are played are defined as neighborhoods of the nodes. In this way each 
agent participates in the group defined by its neighbors (and itself) and in the groups defined as the neighborhoods 
of all its neighbors. For many different types of lattices and random networks, as well as different implementations of evolutionary dynamics, it is found that the kind of assortment introduced by  the network structure promotes cooperation in the sense that full cooperation can be obtained for values of the enhancement factor $r<N$, {\em i.e.}, below the predictions of the well-mixed case. Moreover, the structure of the network itself seems to have an important influence for the promotion of cooperation so that scale-free (SF) networks, for which the probability $P(k)$ that a node has $k$ neighbors follows a power-law $P(k)\sim k^{-\gamma}$, show the smallest value of the enhancement factor $r$ for which cooperation  is obtained \cite{Santos08}.

Another step further in the implementation of the evolutionary dynamics of the PGG have been recently proposed in  \cite{GardenesPGG_11,GardenesPGG2011,Pena12}. In these works, the actual structure of social groups is taken into account, thus getting rid off the above assumption considering that groups are formed by each node together with its neighbors. In fact, this assumption is somewhat unrealistic as it does not reflect the social structure of groups as revealed,  for example, by real collaboration networks \cite{Newman01a,Newman01b,Newman04}. 
To incorporate the group structure into the formulation of the PGG in \cite{GardenesPGG_11,GardenesPGG2011,Pena12} the authors make use of bipartite graphs. In a bipartite graph there are two different types of  nodes representing respectively, in the case of the PGG, agents and groups. In this way, the edges are restricted to connect nodes of  a different type, so that an agent is linked to the groups it participates in, and correspondingly a  group is linked to its forming members. It is important to stress that, unlike the previous networked settings, the size of the groups and  the agent's degree ({\em i.e.}, the number of groups it belongs to) are here disentangled so that an agent $i$ interacting with $k_i$ neighbors is not forced to interact with all of them simultaneously in a single group. Therefore, the bipartite formulation allows to study, {\em e.g.}, different types of probability densities for the number of connections of the agents and for the number of elements of the groups. In particular, in \cite{GardenesPGG_11,GardenesPGG2011} it is shown that the use of realistic patterns of connections for agents and groups leads to an increase of the cooperation with respect of the aforementioned case of scale-free networks.
In brief, the recent works on the evolutionary dynamics of the PGG (see \cite{GardenesPGG2013} for a recent comprehensive review) have tried to explain the emergence of cooperation by
incorporating the actual structure of group interactions relying on real data about social and collaboration networks.

In this paper, following the avenue of encoding the group interaction of the PGG in a bipartite network, we aim at studying the role that information about the payoffs of the agents has on the dissemination of cooperative traits in the PGG. A general underlying assumption in the previously reported studies on the evolutionary dynamics of PGG on bipartite networks is that the fitness of individuals is the simple sum of the payoffs received from  all the games in which they participate. Thus when each agent $i$ compares their fitness with another  agent $j$ in the same group, $i$ is aware of the benefits obtained by $j$ in all the groups it participates, regardless of whether or not $i$ also participates in these groups. In other words, it is assumed not only that a perfect information is available, but also that the whole information from all the groups $j$ participates in is equally relevant for the decission of agent $i$. 
To analyze the role that perfect or partial information about the fitness of individuals has on their decision and the long-term success of cooperation, we introduce  a parameter $0 < \alpha < 1$ to quantify the relevance that payoffs obtained by an individual in other  groups has in its effective fitness relative to the group in which the strategy update takes place. Our results show that perfect information is essential for the development of cooperative traits in structured populations.

The paper is structure as follows. In the next section \ref{TheModel} we introduce in detail the bipartite structure of the population, the specifics of the evolutionary dynamics employed, and the computational details that we have  used in the numerical simulations. In section \ref{Results} we show and discuss the results obtained. Finally, we present some conclusions in section \ref{Conclusions}.

\section{The model}
\label{TheModel}

\subsection{The structure of interactions}
As introduced before we consider a population of $N$ agents that interacts within $N_G$ groups as dictated by a bipartite graph. In this way each agent $i$ belongs to $k_i$  ($k_{min} \leq k_i \leq k_{max}$) groups, whereas a group $g$ contains $G_g$ ($G_{min} \leq G_g \leq G_{max}$) agents. We assume  that there is no internal structure inside each group. In order to characterize the interaction patterns between individuals and groups it is common to consider the degree distribution of the agents $P(k)$ and the size distribution $P(G)$ of the groups. By fixing $P(k)$ and $P(G)$ one can construct random bipartite graphs so that each particular realization of the graph is specified by a $(N_G \times N)$ membership matrix  ${\bf A}$ defined as $A_{g,i}=1$ whenever $i \in g$, or $A_{g,i}=0$  otherwise. Given ${\bf A}$ it is easy to express the number of groups, $k_i$, the  agent $i$ takes part in as:
\begin{equation}
k_i = \sum_{g=1}^{N_G} a_{g,i} \;\;\; (i= 1, ..., N)\;,
\label{agentdegree}
\end{equation}
and the number of participants, $G_g$, contained in group $g$, as:
\begin{equation}
G_g = \sum_{i=1}^{N} a_{g,i} \;\;\; (g= 1, ..., N_g)\;.
\label{groupsize}
\end{equation}

Let us note that, in finite populations, not all the combinations of $P(k)$ and $P(G)$ are compatible. In particular, the total number of links in the bipartite graph is the same no matter how they are computed considering either the agents or the groups. This latter closure condition yields the following identity:
\begin{equation}
\sum_{i=1}^{N} k_i = \sum_{g=1}^{N_G} G_g \;\;\; \Rightarrow N \sum_k P(k) k = N_G\sum_G P(G) G\;,
\label{compatibility}
\end{equation}
that automatically sets a compatibility condition for the degree and group size distribution densities. This fact introduces a constraint in (at least) one of the two distributions $P(k)$ and $P(G)$ as we will see below.

\subsection{The Public Goods Game in bipartite graph}
Now we go over the nature of agents' interactions as players of a PGG. Each group is 
the scenario of a PGG, and all the groups share the same rules for the game, which 
implies that there is a common enhancement factor $r$ for all PGGs. In this paper an agent $i$ 
at time $t$ adopts the same action (to cooperate C or to free ride D) on all the different $k_i$ 
groups it belongs to. At each dynamical step $t$, we characterize the internal state of each 
agent $i$ by the strategic variable $s_{i}(t)$, which can take two possible values: $1$ if $i$ 
cooperates, or $0$ when the agent free rides. This strategic variable, together with the membership 
matrix ${\bf A}$ that fixes the social structure, completely specifies the instantaneous microscopic state 
of the population. According to this, the fraction of cooperators in group $g$ ($g= 1, ..., N_G$) 
at time $t$ is given by:
\begin{equation}
x_g(t)= \frac{1}{G_g} \sum_{i \in g} s_{i}(t) \;,
\end{equation}
whereas its average value over the groups,
\begin{equation}
\bar{x}(t)= \frac{1}{N_G} \sum_g x_g(t) \;,
\label{groupcoop}
\end{equation}
defines the instantaneous average group cooperation. In general, the value $\bar{x}(t)$ is different from the fraction of cooperators at time $t$:
\begin{equation}
c(t) = \frac{1}{N} \sum_{i=1}^N s_i(t)\;.
\label{fraccoop}
\end{equation}
On the other hand, the value:
\begin{equation}
c_a(t)= \frac{\sum_{i =1}^N k_i s_i(t)}{\sum_{i =1}^N k_i} = \frac{\sum_{g=1}^{N_G}G_g x_g(t)}{\sum_{g=1}^{N_G}G_g}\;,
\label{coopaction}
\end{equation}
represents the fraction of cooperative actions averaged over all the agents and PGGs in which 
they participate. Note that for general $P(k)$ and $P(G)$ compatible densities, the three averages 
$\bar{x}(t)$, $c(t)$, and $c_a(t)$ need not be equal, the reason being that the contribution of each 
particular agent might be weighted differently.

Here we will consider two possible formulations for assigning the value of the investment made 
in each of the PGGs in which an agent participates. On one hand, we study a fixed cost per 
game (FCG) formulation so that each cooperator $i$ invests a fixed cost $c_i=\zeta=1$ in each 
of the $k_i$ games it participates in. On the other hand, we will also consider the formulation of 
fixed cost per individual (FCI) according to which, a cooperator $i$ equally distributes its total 
investment $\zeta$ among all its groups, so that the agent invests $c_i=\zeta/k_i=1/k_i$ in each PGG. In 
both cases, the total contribution of all cooperators in a group is multiplied by an enhancement 
factor $r$ and the result is equally distributed between all the $G_g$ members of the group.

According to the FCG formulation, each player $i$ receives a payoff:
\begin{equation}
\pi_{g,i}(t) = rx_g(t)-s_i(t)\;,
\label{benefitFCG}
\end{equation}
from its participation in a group $g$ and, otherwise, according to the FCI formulation it receives:
\begin{equation}
\pi_{g,i}(t) = \frac{r \sum_{j \in g} s_j(t)/k_j}{G_g}-s_i(t)k_i^{-1}\;.
\label{benefitFCI}
\end{equation}


Given an agent $i$ and a group $g$ of which it is a member, the effective fitness of 
agent $i$ relative to group $g$ is defined as
\begin{equation}
\pi^{eff}_{g,i}(t) = \pi_{g,i}(t) + \alpha \sum_{g^{\prime}=1}^{N_G} (A_{g^{\prime},i}-\delta_{g^{\prime},g})\cdot\pi_{g^{\prime},i}(t)  \;,
\end{equation}
where $\delta_{g^{\prime},g}=1$ when $g^{\prime}=g$ and $\delta_{g^{\prime},g}=0$ otherwise. The parameter $\alpha\in [0,1]$ in the above expression quantifies the amount of information shared between groups.  In particular, the higher $\alpha$, the more information is shared, so that $\alpha=0$ corresponds to the extreme case in which each agent's effective fitness relative to a given group uses only  
the information on payoffs in this group. On the other hand, for $\alpha=1$ agents have full information about payoffs from other groups.

\subsection{The evolutionary Dynamics}
Once all the PGGs have been played in the groups and the corresponding payoffs have been collected by the individuals the evolutionary dynamics takes place. In this work we consider the {\em Fermi Rule} as the strategic update framework for the evolutionary dynamics 
\cite{SzaboPRE1998,TraulsenPRE2006}. In this way, at each time step $t$, each agent $i$  chooses a random group $g$ among all its groups and a random partner, say $j$, of the chosen group $g$, and compares their effective fitness relative to group $g$. Agent $i$ will 
imitate the action of $j$, $(s_i(t+1)=s_j(t))$, with a probability given by a Fermi function of the effective fitness difference:
\begin{equation}
P_{i\rightarrow j}=\frac{1}{1+exp(-\beta (\pi^{eff}_{g,j}(t)-\pi^{eff}_{g,i}(t)))}   \;,
\label{Fermi}
\end{equation}
where $\beta$ is a constant often interpreted as the inverse temperature of the system, \textit{i.e.}, the higher temperature, the more random is imitation and thus, the more smoothly dependent on the differences of effective fitness (payoffs).

\section{Results}
\label{Results}

In order to study the influence of the information exchange between groups in the evolution of cooperation we have considered two different scenarios. In the first one, all the groups have the same size $G_g=G$, while the connectivity of the agents (\textit{i.e.}, the number of groups to which they belong) varies from one to another being distributed according to a power law, {\em i.e.}, the probability for an agent $i$ to participate in $k_i$ groups is given by $P(k_i)=Ck_i^{-\gamma}$, for certain constants $C, \gamma$. In the second 
scenario, we have fixed the agents' connectivity $k_i=k$ and distributed the group size according to a power law, being the probability $P(G_g)$ for a given group $g$ to have $G_g$ members: $P(G_g)=C^{\prime}G_g^{-\gamma^{\prime}}$. 

The two scenarios described above, along with the FCG and FCI formulations of PGG, provide four combinations that we have separately studied through numerical simulations. In the first scenario, the connectivity of the agents is given by:
\begin{equation}
P(k_i)= C k_i^{-\gamma} =\left( \sum_{k=k_{min}}^{k_{max}}k^{-\gamma}\right)^{-1} k_i^{-\gamma}   \;, 
\label{distribution1}
\end{equation}
where $k_{min}$ and $k_{max}$ are, respectively, the minimum and maximum possible 
connectivities. In addition, the condition (\ref{compatibility}), which here takes the form 
$\langle k\rangle N=G N_G$, implies:
\begin{equation}
C=\frac{GN_G}{N\sum_{k=k_{min}}^{k_{max}}k^{1-\gamma}}\;
\end{equation}
and therefore:
\begin{equation}
\frac{\sum_{k=k_{min}}^{k_{max}}k^{-\gamma}}{\sum_{k=k_{min}}^{k_{max}}k^{1-\gamma}}=\frac{N}{GN_G} \;,
\label{normalization.Scenario1}
\end{equation}
that provides the value of exponent $\gamma$ once the network parameters
$N, N_G, G, k_{min}$ and $k_{max}$ are fixed. Let us note that in this first scenario
the instantaneous group cooperation, equation (\ref{groupcoop}), coincides 
with the fraction of cooperative actions, equation (\ref{coopaction}), $\bar{x}(t) =c_a(t)\;$.

Alternatively, in the second scenario the size of the groups is given by:
\begin{equation}
P(G_g)= C^{\prime} G_g^{-\gamma^{\prime}} =\left( \sum_{G=G_{min}}^{G_{max}}G^{-\gamma^{\prime}}\right)^{-1} G_g^{-\gamma^{\prime}}   \;, 
\label{distribution2}
\end{equation}
where $G_{min}$ and $G_{max}$ are, respectively, the minimum and maximum 
possible sizes. Equation (\ref{compatibility}) introduces a constraint for the group 
size distribution density $P(G_g)$, and the analogous to expression 
(\ref{normalization.Scenario1}) takes the form:
\begin{equation}
\frac{\sum_{G=G_{min}}^{G_{max}}G^{-\gamma^{\prime}}}{\sum_{G=G_{min}}^{G_{max}}G^{1-\gamma^{\prime}}}=\frac{N_G}{kN} \;.
\label{normalization.Scenario2}
\end{equation}
Finally, let us note that in the second scenario the fraction of cooperators, equation (\ref{fraccoop}), 
coincides with the fraction of cooperative actions, equation (\ref{coopaction}), $c(t) =c_a(t)\;$.

\begin{figure}
\begin{center}
\epsfig{file=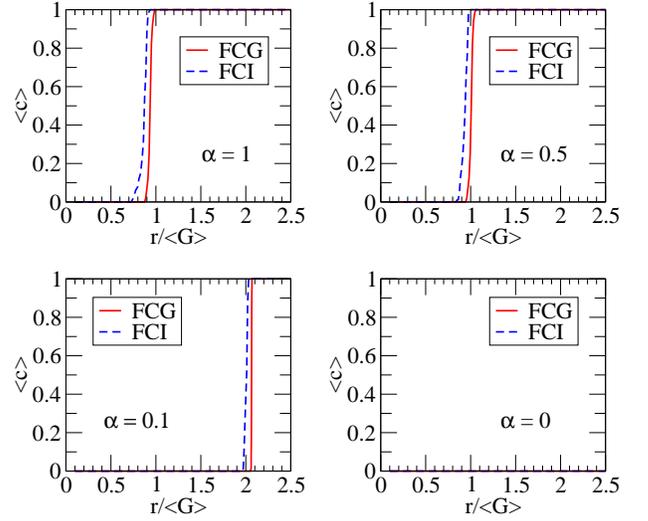,width=0.38\textwidth,angle=-90,clip=1}
\end{center}
\caption{(Color online) Average fraction of cooperators $\langle c\rangle$, as a function of the normalized enhacement factor
$r/\langle G\rangle$, for a SF distribution of the membership frequency $k_i$ (\textit{i.e.}, number of groups to which an agent belongs), and a fixed group size $G_g=10$. Different panels correspond to different values
of the parameter $\alpha$ that models the inter-groups
information exchange: $\alpha=1$ corresponds to the case in which agents have full information about their partners' payoffs,
while for $\alpha=0$ agents only have knowledge of payments from their games. Solid red lines correspond to the FCG formulation of
the PGG, while dashed blue lines correspond to the FCI version. Results are averaged over $10^2$ realizations (and
different membership matrices) for each value of $r/\langle G\rangle$. Other parameter
values are $N=N_G=10^3$, $\beta=0.1$, $\zeta=1$.}
\label{fig.c.Vs.r.fixedG}
\end{figure}

We simulated the evolutionary dynamics of PGG, starting from an initial condition 
according to which the fraction of cooperators approximately equals the number of 
free-riders and both strategists are randomly distributed. For each value of the 
normalized enhancement factor $r/\langle G \rangle$, we iterate a large number of 
rounds ($10^5$) and measure the average fraction $c$ of cooperators averaged 
over a time window of $10^4$ additional rounds. In addition, we reported the values 
once overaged over $10^2$ different networks -membership matrices- and initial 
conditions. The number of groups and agents were both fixed to $N_G=N=10^3$. 
After the random assignation of the agents' connectivities according to the distribution 
given by formula (\ref{distribution1}) for the first scenario (or, respectively, the group 
sizes following the distribution given by (\ref{distribution2}) for the second scenario) 
we implemented the network following a configurational model. The minimum size  
was fixed to $G_{min}=2$ in order to avoid one-player games; likewise, the minimum 
connectivity was also fixed to $k_{min}=2$ to allow comparisons between scenarios. 
Maximum size and connectivity were fixed to $G_{max}=k_{max}=32$, implying 
$C=C^{\prime}\simeq 0.3399$ and $\gamma=\gamma^{\prime}\simeq 1.020$ (these values were numerically 
calculated). 

Regarding the first scenario, in which we have fixed the size of the groups to $G=10$ 
and distributed the connectivity of the agents according to a power law, Fig.~\ref{fig.c.Vs.r.fixedG} 
shows the average fraction $\langle c \rangle$ of cooperators \cite{note} as a function of 
the normalized enhancement factor $r/G$ for the PGG, according to the FCG (solid red lines) 
and to the FCI (dashed blue lines) formulations, for a value of the parameter $\beta=0.1$. Each panel in 
Fig.~\ref{fig.c.Vs.r.fixedG} corresponds to a different value of the inter-groups information exchange parameter 
$\alpha$, from $\alpha=1$ (left top panel) which corresponds to the case in which perfect 
global information about partners' payoffs is available to the agents, to $\alpha=0$ (right 
bottom panel) where effective fitness of agents only depends on local information. As it
can be observed, there exists a critical value $r_c/G$ at which a phase transition takes 
place. The main finding is that the value $r_c/G$ increases with decreasing $\alpha$, 
which means that global information plays a positive role in the evolution of cooperation. 
In fact, while the transition to cooperative states take place at $r_c/G\simeq 0.9$ for 
the FCG formulation (\textit{resp.}, $r_c/G\simeq 0.7$ for FCI) when $\alpha=1$, this 
critical value increases as we decrease the parameter $ \alpha$ and, finally, no transition 
occurs for $\alpha=0$.

The absence of a transition to a cooperative regime for $\alpha = 0$ (when the effective 
fitness of an individual relative to a group is just its payoff from this group) 
can be easily understood: the benefit obtained by an agent from a single group $\pi_{g,i}$, 
see equations (\ref{benefitFCG}) and (\ref{benefitFCI}), is always higher for a free rider
than for a cooperator in the same group, irrespective of the formulation (FCG or FCI) 
used, and of the value of the normalized enhancement factor $r/G$. Then, for strategy 
updating rules based on fitness differences, the bias in the imitation probability against 
cooperators forces always the evolutionary fixation of free riders in the long term, and 
the transition cannot take place. We thus clearly see that a value of $\alpha > 0$ is indeed 
a prerequisite for the eventual existence of a cooperative regime. Once this condition is 
satisfied, the role of those individuals that participate in many games (hubs) in promoting the 
cooperative transition is, in the scenario of figure \ref{fig.c.Vs.r.fixedG}, an important issue, 
as in other different previously studied settings for the evolutionary dynamics of PGG 
\cite{Santos08,GardenesPGG2011} as well as of two-person games \cite{Santos05,Santos06,Gardenes2007,Floria}. 
Briefly said, hubs (either cooperators or free riders) collect very high payoffs and are neighbors 
of many agents, and so they are easily imitated by individuals in their (many) groups. However, 
while the imitation of free rider hubs decreases their future benefits (a source of future 
instability), just the opposite happens to cooperator hubs when imitated by partners (stability increases).  

On the other hand, comparing the results obtained with both formulations, one observes 
that the FCI version presents lower values of $r_c$ than those corresponding to FCG, 
which means that the FCI formulation promotes better the convergence to 
cooperative states for any value of $\alpha$ that enables the transition. This is due to 
the fact that in the FCG formulation cooperators pay according to their connectivity 
(the more games, the higher the cost), while in the FCI version cooperators pay a fixed 
cost regardless of the number of PGG in which they participate. This independence 
of the cost of cooperation (investment) on connectivity provides cooperators, and specially cooperator hubs, 
higher payoffs in the FCI formulation, and the survival of cooperation is consequently 
enhanced \cite{GardenesPGG2011}. 

\begin{figure}
\begin{center}
\epsfig{file=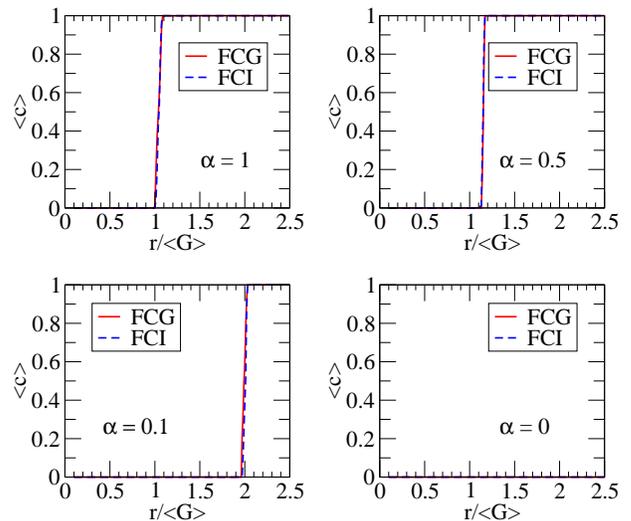,width=0.38\textwidth,angle=-90,clip=1}
\end{center}
\caption{(Color online) Average fraction of cooperators $\langle c\rangle$, as a function of the normalized enhacement factor
$r/\langle G\rangle$, when the size of the groups $G_g$ are distributed according to a power-law and all the agents
belong to the same number of groups $k_i=10$. Different panels correspond to different values of the inter-groups information exchange
ratio, from $\alpha=1$ (maximum) to $\alpha=0$ (minimum). Solid lines red are for the FCG formulation, while dashed blue lines are for the FCI version. Results are averaged over $10^2$ realizations performed in different networks. Other parameter are equal as those used in Fig.~\ref{fig.c.Vs.r.fixedG}.}
\label{fig.c.Vs.r.fixedK}
\end{figure}

Figure \ref{fig.c.Vs.r.fixedK} represents the results for the second scenario, in which all the 
agents have the same connectivity $k=10$ (\textit{i.e.}, all the agents participate in exactly 
10 games) and the group size distribution $P(G_g)$ follows a power law 
$P(G_g)=C^{\prime}G_g^{-\gamma^{\prime}}$. The four panels show the asymptotic average fraction $\langle c \rangle$ of 
cooperators versus the normalized enhancement factor $r/\langle G\rangle$ for different 
values of the parameter $\alpha$, from $\alpha=1$ (left top panel, full payoff information 
available) to $\alpha=0$ (right bottom panel, pay-off information restricted to the focal 
group). Note that the argument given above showing that no cooperative regime exists for 
$\alpha = 0$, is also valid in this second scenario. For $\alpha>0$ however, as there are 
no hubs here, a different mechanism for the possibility of a cooperative regime must be 
invoked. If one thinks of the situation in which all the groups have the same fraction $x_g<1$ of 
cooperators, one easily realizes that every free rider earns more effective fitness than 
any cooperator, which will, in the long term suppress cooperation. In other words, 
fluctuations of $x_g$ among the different groups seems to 
be needed for the evolutionary success of cooperation. Let us remember that in the initial 
condition the strategies are randomly distributed among the agents, and that the scale-free 
distribution of group sizes makes small groups abundant in this scenario. It is precisely 
among small groups where the likelihood of an initial fraction 1 of cooperators is largest. 
The high payoffs received by these cooperators from the small-sized fully cooperative 
groups makes plausible, provided $\alpha >0$, that they could both resist invasion from free 
riders and spread cooperation at updating trials in the larger groups they participate in.

In Fig.~\ref{fig.c.Vs.r.fixedK}, almost no difference is observed in the results obtained for 
both (FCG and FCI) formulations, a feature that (at a first sight) could be explained 
because now all the players are involved in the same number of games, and so the relative 
benefits of strategists do not vary from one to the other formulation. However, see below 
(at the end of this section) for a more precise argument that takes into account the role of $\beta$ 
and reveals the weakness of this first explanation. 
On the other hand, by comparing Figs.~\ref{fig.c.Vs.r.fixedG} and \ref{fig.c.Vs.r.fixedK},
it can be seen that for high and intermediate values ​​of $\alpha$
the critical value of the enhancement factor $r_c$ in the second scenario with
non-uniform group sizes is higher than on the first
one with heterogeneous connectivity, which shows the effectiveness of the role of hubs
in promoting cooperation provided that players are aware of their partners' payments from 
outside the focal group.

\begin{figure}
\begin{center}
\epsfig{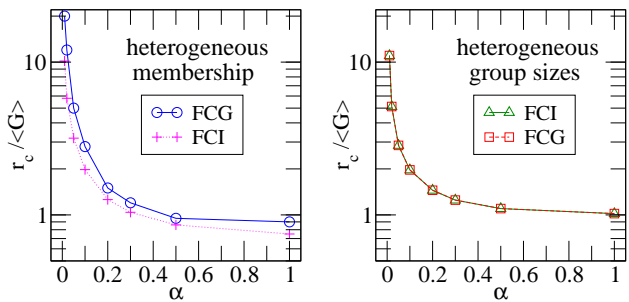}
\end{center}
\caption{(Color online) Critical value $r_c/\langle G\rangle$ of the normalized enhacement factor
as a function of the inter-groups information exchange ratio $\alpha$ for the FCG and FCI formulations of the PGG. Left panel corresponds to 
a SF distribution of the number of groups to which an agent belongs, $k_i$, and a fixed group size $G_g=10$. Right
panels corresponds to the case in which the group size distribution ($P(G_g)$) follows a power-law while all the agents
belong to the same number of groups $k_i=10$. Other parameter are equal as those used in Fig.~\ref{fig.c.Vs.r.fixedG}.}
\label{fig.critical.c.Vs.r}
\end{figure}

The influence of the inter-groups information exchange on the evolution of cooperation is more
clearly shown in Fig.~\ref{fig.critical.c.Vs.r} where we show the critical value 
$r_c/\langle G\rangle$ of the normalized enhancement factor versus the value of the inter-groups 
information exchange parameter $\alpha$, for FCG and FCI formulations and both scenarios
considered: a fixed group size $G=10$ in the left panel and a fixed connectivity $k=10$ in 
the right panel. On the one hand, it can be noticed the abrupt nature of the dependence: while 
for high values of $\alpha$ the curve is smooth and $r_c/\langle G\rangle$ does not present 
a strong dependence on $\alpha$, for $\alpha \simeq 0$ a small decrease of $\alpha$ implies 
a very large increase of $r_c$. This fact indicates that only a small information exchange is enough 
to trigger the emergence of cooperation. On the other hand, as it has been previously mentioned, 
when the distribution of the number of groups, $k_i$, to which an agent $i$ belongs is 
heterogeneous (left panel), the FCI formulation presents lower values of the critical value $r_c$ 
for any value of $\alpha>0$ than the FCG version and therefore a higher tendency to cooperative 
states; on the contrary, when the group size distribution $P(G_g)$ follows a power-law and 
all the agents participate in the same number of games, $k_i=10$, the FCG and FCI curves 
match which indicates that the critical value $r_c$ is robust against the choice of formulation.

All the previous shown results have been calculated for a particular value of the parameter 
$\beta=0.1$, where $1/\beta$ mimics somehow the temperature of the system: the higher 
temperature the more random is imitation and thus less dependent on the difference of effective fitness. 
At variance with other updating rules, as {\em e.g.}, discrete replicator or imitate the best, whose dynamics very often reaches stationary states of strategic coexistence, the "trembling hand" mechanism that the Fermi rule incorporates through 
$\beta$ has the effect that wherever it has been used, evolution seems to end up in 
absorbing mono-strategic states. This is in fact the case in the first scenario, so that 
those situations in Fig.~\ref{fig.c.Vs.r.fixedG} where $0 < \langle c \rangle < 1$ means 
that the fraction of realizations ending up in a fully cooperative absorbing 
stat is $\langle c \rangle$. However, this is no longer the case when $P(G_g)$ follows a 
power-law and all the agents participate in $k=10$ games. In other words, 
a value $0 < \langle c \rangle < 1$ of the average fraction of cooperators in Fig.~\ref{fig.c.Vs.r.fixedK}, corresponds to states of true strategic coexistence in the realizations.

\begin{figure}
\begin{center}
\epsfig{file=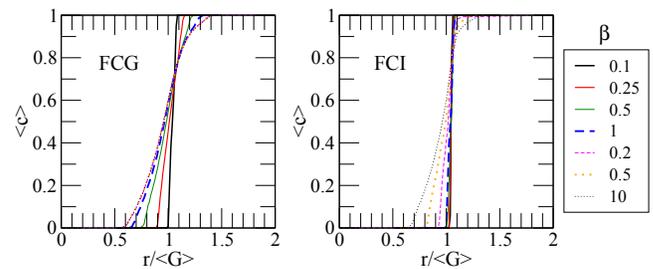,width=0.47\textwidth,angle=0,clip=1}
\end{center}
\caption{(Color online) Average fraction of cooperators $\langle c\rangle$ versus the normalized enhacement factor $r/\langle G\rangle$ for different values of the parameter $\beta$ 
corresponding to the inverse temperature of the system (\textit{i.e.}, the lower
$\beta$, the more random is imitation), when the group
size distribution $P(G_g)$ follows a power-law and all the agents
belong to the same number of groups $k_i=10$. Left panel corresponds to the FCG formulation of
the PGG, while right panel corresponds to the FCI one. Results are averaged over $10^2$ realizations performed in different networks. Other parameter are equal as those used in Fig.~\ref{fig.c.Vs.r.fixedG}.}
\label{fig.c.Vs.r.beta}
\end{figure}

We show in Fig.~\ref{fig.c.Vs.r.beta} the asymptotic average fraction of cooperators $\langle c \rangle$ as a 
function of $r/\langle G\rangle$ for different values of 
the parameter $\beta$, for the second scenario, full information ($\alpha =1$), and for both formulations: 
left panel corresponds to the FCG formulation while right panel corresponds to the FCI 
version. As it can be seen, while for high values of the temperature (\textit{i.e.}, low values 
of $\beta$) the transition is abrupt, with a sharp shift from the uncooperative state $c=0$ 
to the fully cooperative state $c=1$, for low values of the temperature the transition is 
smooth. This is due to the fact that, according to equation (\ref{Fermi}), at low temperatures 
the imitation probability as a function of the difference of effective fitness is very close to 
a step function (no imitation unless other's fitness is higher), and therefore the probability 
to get stuck in metastable non-absorbing states with $0<c<1$ is higher than for high 
temperature, where the likelihood for players to imitate partners with lower 
benefits is greater, so allowing the dynamics to explore wider regions of the space of 
configurations and to find a way out to an absorbing state. 

The comparison of panels in Fig.~\ref{fig.c.Vs.r.beta} reveals a remarkable similarity 
of both set of curves, despite the fact that for high $\beta$ individual curves clearly differs
from panel to panel. This observation has a crystal clear explanation from the analysis of 
equations (\ref{benefitFCG}) to (\ref{Fermi}). In the second scenario, where all agents have 
the same connectivity, for every given strategic configuration, and 
irrespective of the value of $\alpha$, the effective fitness of all the agents are simply 
multiplied by a constant factor $k$ when changing from the FCI formulation to the FCG one. 
Thus, from equation (\ref{Fermi}), all the transition probabilities between configurations for 
the FCI formulation at a value $\beta$ are exactly those for the FCG one at a value $\beta/k$. 
Note that this argument further modifies 
the explanation given above for the almost coincidence observed in the behavior of the curves 
$\langle c \rangle$ vs. $r/\langle G \rangle$ for FCI and FCG formulations in Fig.\ref{fig.c.Vs.r.fixedK}, in the 
sense that this coincidence is due to the small value of $\beta=0.1$ used there; larger values of 
$\beta$ would have revealed stronger differences nearby the transition, that nonetheless do not affect the location of 
the critical values $r_c/\langle G \rangle$.

\section{Conclusions.}
\label{Conclusions}
It is well known that for the study of the evolution of cooperation in most real
systems is of utmost importance to deal with N-agents interactions as well as
to take into account the details of the group structure in which interactions take place. Recently, several mechanisms have been suggested for explaining the onset of cooperative behavior in the PGG by considering bipartite graphs. In this kind of graphs it is possible to reproduce the assortment of individuals in groups with distribution of sizes according to that observed in real systems. In this paper, by exploiting the bipartite framework, we have considered the role that information exchange between groups has on the emergence of cooperation in the PGG. In particular, we have studied how the knowledge of the payoff collected by an individual in all the groups it belongs to affect the decision of those individuals interacting with it in a single group. 

Our results point out that cross-information plays a fundamental and positive role in the evolution of cooperation. Moreover, by tuning a parameter quantifying  the interchange of information between groups we have observed that the influence of this latter ingredient is highly nonlinear: a small amount of information exchange is enough to promote cooperation to the levels previously observed in bipartite graphs. Let us note that the limit in which no information interchange is allowed is similar to the well mixed scenario, in which cooperation is not observed unless the enhancement factor $r$ is larger than the typical size of the groups. 

Furthermore, by comparing setups with either a heterogenous distribution for the connectivity of the agents (combined with a homogeneous distribution for the size of the groups) or a heterogeneous distribution for the sizes of groups (combined with a homogeneous distribution for the agents connectivities), we have shown that systems with heterogeneous connectivities for the agents converge to cooperative states for lower values of the enhancement factor $r$, thus displaying the role of hubs in promoting cooperation in PGGs. In addition, within this latter setup, the FCI framework (in which cooperators pay a fixed cost regardless the number of groups they belong to) presents lower values of the critical enhancement factor $r_c$ than those corresponding to FCG
setting (in which the cost of a cooperator is proportional to the number of groups it belongs to) due to the higher benefits of cooperator hubs in the FCI setup. Nevertheless, when the group size distribution is heterogeneous and all the agents
belong to the same number of groups, both the FCG and the FCI version of the PGG show roughly the same behavior. 

Summarizing, our results show that the exchange of information between groups enhances the cooperation in interconnected populations, highlighting the idea that a collaborative attitude in systems composed of many local groups can be fostered by facilitating the ways of communication beyond single work teams.

\acknowledgments
We acknowledge financial support from Spanish MINECO under projects FIS2011-25167 and FIS2012-38266-C02-01 and from the Comunidad de Arag\'on (Grupo FENOL). JGG is supported by the Spanish MINECO through the Ramon y Cajal program. YM acknowledges partial financial funding from EU FET Proactive project MULTIPLEX (contract no. 317532).


\begin{thebibliography}{99}
\bibitem{note} It turns out that the average over realizations of both, the fraction of cooperators 
and the fraction of cooperative actions, numerically coincides for this scenario. The reason for 
this coincidence is that all the realizations converge to one or the other of the two absorbing states, 
so that these averages simply are the fraction of realizations ending at full cooperation.
\end{thebibliography}
\end{document}